# Laser driven melt pool resonances through dynamically oscillating energy inputs

*Marco Rupp[1,2], Karen Schwarzkopf [3,4], Markus Döring[3,4], Shuichiro Hayashi[2], Michael Schmidt[3,4], Craig B. Arnold[1,2,*]*

1-Department of Mechanical and Aerospace Engineering, Princeton University, Princeton, 08544, NJ, USA.

2-Princeton Materials Institute, Princeton University, Princeton, 08544, NJ, USA

3- Institute of Photonic Technologies, Friedrich-Alexander-Universität Erlangen-Nürnberg, Erlangen, 91052, Germany

4- Erlangen Graduate School in Advanced Optical Technologies, Friedrich-Alexander-Universität Erlangen-Nürnberg, 91052 Erlangen, Germany

*cbarnold @princeton.edu

Abstract

Spatially selective melting of metal materials by laser irradiation allows for the precise welding as well as the 3D printing of complex metal parts. However, the simple scanning of a conventional Gaussian beam typically results in a melt track with randomly distributed surface features due to the complex and dynamic behavior of the melt pool. In this study, the implications of utilizing a dynamically oscillating energy input on driving melt track fluctuations is investigated. Specifically, the laser intensity and/or intensity distribution is sinusoidally modulated at different scan speeds, and the effect of modulation frequency on the resulting surface features of the melt track is examined. The formation of periodically oriented surface features indicates an evident frequency coupling between the melt pool and the modulation frequency. Moreover, such a frequency coupling becomes most prominent under a specific modulation frequency, suggesting resonant behavior. The insights provided in this study will enable the development of novel methods, allowing for the control and/or mitigation of inherent fluctuations in the melt pool through laser-driven resonances.

1. Introduction

Laser material processing offers locally confined energy inputs, enabling the spatially selective heating of various materials, including ceramics, polymers, and metals.[1–4] Particularly in the context of metals, by harnessing the high spatial resolution and digital customizability, various innovative manufacturing techniques, namely laser welding and laser powder bed fusion, have been developed to further advance the aerospace, automotive, and biomedical industries.[5–7] However, the rapid heating processes involved in laser melting can result in a multitude of complexities within the melt pool, which lead to the formation of various defects.[8] Such defects are potential precursors of structural failure and significantly reduce the reliability of the processed part.[9] Therefore, shedding light on the various complexities in the melt pool is a crucial step towards the development of a technique capable of the defect-free laser melting of metals.[10,11]

The intensity of the laser beam is related to the degree of material melting, and is a key parameter that determines the state of the melt pool.[12] By modifying the spatial distribution of laser intensity, different properties of the melt pool can be altered, such as melt pool geometries, fluid instabilities, and melting/solidification behaviors.[13,14] Therefore, beam shaping has been increasingly implemented for melt pool control in laser melting , allowing for the reduction of residual stresses and porosity formation, as well as tuning of metal grain sizes.[15,16] For example, Nie et al. have indicated that the expansion of the spot size of an equiaxial Gaussian beam via beam defocusing can increase the melt pool dimensions, resulting in crack-free melt tracks owing to prolonged solidification.[17] Furthermore, Roehling et al. have indicated that by changing the beam profile from Gaussian to elliptical, the thermal gradient across the melt pool can be lowered, resulting in larger grains.[18] Recently, Zhang et al. have shown that fluid instabilities can be intentionally driven within the melt pool by super-positioning two independent equiaxial Gaussian beams with a slight spatial offset to form ordered periodic metal structures with high repeatability.[19]

The state of the melt pool is inherently dynamic due to the complex and ever-changing interactions between the laser beam and the material.[20] Therefore, dynamic techniques which temporally modulates the parameters of the laser beam during melting have been developed to address this challenge. Ning et al. have shown that by modulating the laser intensity, the welding efficiency of highly reflective metals can be improved.[21] They discuss that such a modulation alters the transient geometry of the melt pool, enhancing the absorption of laser energy and stabilizing the melt process. Moreover, Girerd et al. have investigated the advantages of complex dynamic spatial modulation schemes, commonly referred to as beam oscillation or wobbling[22,23], on the laser welding of metals.[24] The rapid scanning of the laser beam with a periodically oscillating pattern, such as circular or Lissajous, induces gradual thermal accumulation, allowing for precise control of thermal gradients and phases of the resulting metal grains. The results from Ning and Girerd highlight the effectiveness of utilizing dynamic techniques where the parameters are periodically modulated temporally or spatially in controlling the melt pool dynamics. However, to the best of our knowledge, the implications of dynamic techniques in which the parameters are periodically modulated both temporally and spatially have been severely limited.

This study seeks to investigate the effects of dynamically varying both temporal and spatial intensities of the laser beam on the laser melting of metals. Specifically, how sinusoidal modulation of laser intensity and/or intensity distribution while laser scanning influences the surface features of the resulting melt track. Parameters including modulation frequency and scan speed of the laser beam were altered, and the resulting melt track were characterized via confocal microscope observations. Additionally, Fast Fourier transform (FFT) analyses were further conducted to quantitatively assess the formed surface features. Through these characterizations, it was revealed that for a dynamically modulated beam, periodic surface features with a spatial frequency corresponding to the modulation frequency arise on the melt track. Such results indicate the presence of a frequency coupling behavior between the melt pool and modulation frequency. Moreover, such surface features become most prominent under a specific modulation frequency suggesting an underlying resonance. These results provide the implications of dynamically changing both temporal and spatial intensities in driving melt pool resonances.

2. Methods

For all experiments conducted in this study, aluminum alloy metal plates (EN AW-5083) were used as the target metal material.

2.1. Intensity modulation

To probe the effects of intensity modulation, a continuous wave (CW) laser beam setups was utilized. The experiment was performed with a disk laser (TruDisk 8001, Trumpf, Germany) with an operating wavelength of 1030 nm. The integration of a beam shaping module (Canunda, Cailabs, France) into the beam path resulted in ring-core beam profile with a 50:50 power ratio, creating an inner core diameter of 100 μm and an outer ring diameter of 300 μm. The laser scanning speed was set to 1000 mm/s for all experiments using this system. The laser power was modulated sinusoidally from 2640 W to 3960 W (Medium Power: 3300 W) at different frequencies. Fig.1 (a) shows a schematic of the laser beam intensity over one oscillation period.

### 2.2. Intensity distribution modulation

To probe the effect of the intensity distribution and scan speed on the laser processing of metals, one YLR-100 (Maximum output: 100 W) and one YLR-400 (Maximum output: 400 W) laser systems (both IPG Photonics, USA), both generating a CW laser beam with a central wavelength of 1060 nm, were utilized simultaneously. The laser power was set to be the maximum output for the entirety of the experiments. The laser beams, both with a spot size of approximately 100 μm, were focused on the same location of a metal plate and subsequently scanned across the surface using a Focus Shifter digital galvanometer laser scanner system (Raylase GmbH, Germany) at scan speeds of 50, 100, 200, and 400 mm/s. The 400 W laser beam was scanned without oscillation, while the 100 W laser beam was oscillated perpendicular to the scan direction at different frequencies, with a standard amplitude of 30 μm. Fig.1 (b) shows a schematic of the laser beam intensity distribution over one oscillation period.

### 2.3. Characterizations of melt tracks

The height profiles of the melt tracks were optically obtained using confocal microscopy (VR6200 and VK-X3050, Keyence, Japan). To quantitatively characterize the height profiles, FFT analyses were conducted, and the power spectral density ($|X(f)|^2$) was calculated.[25] Detailed observations of the formed surface features were conducted via Scanning Electron Microscopy (SEM, Quanta 200 FEG, Thermo Fisher Scientific, USA).

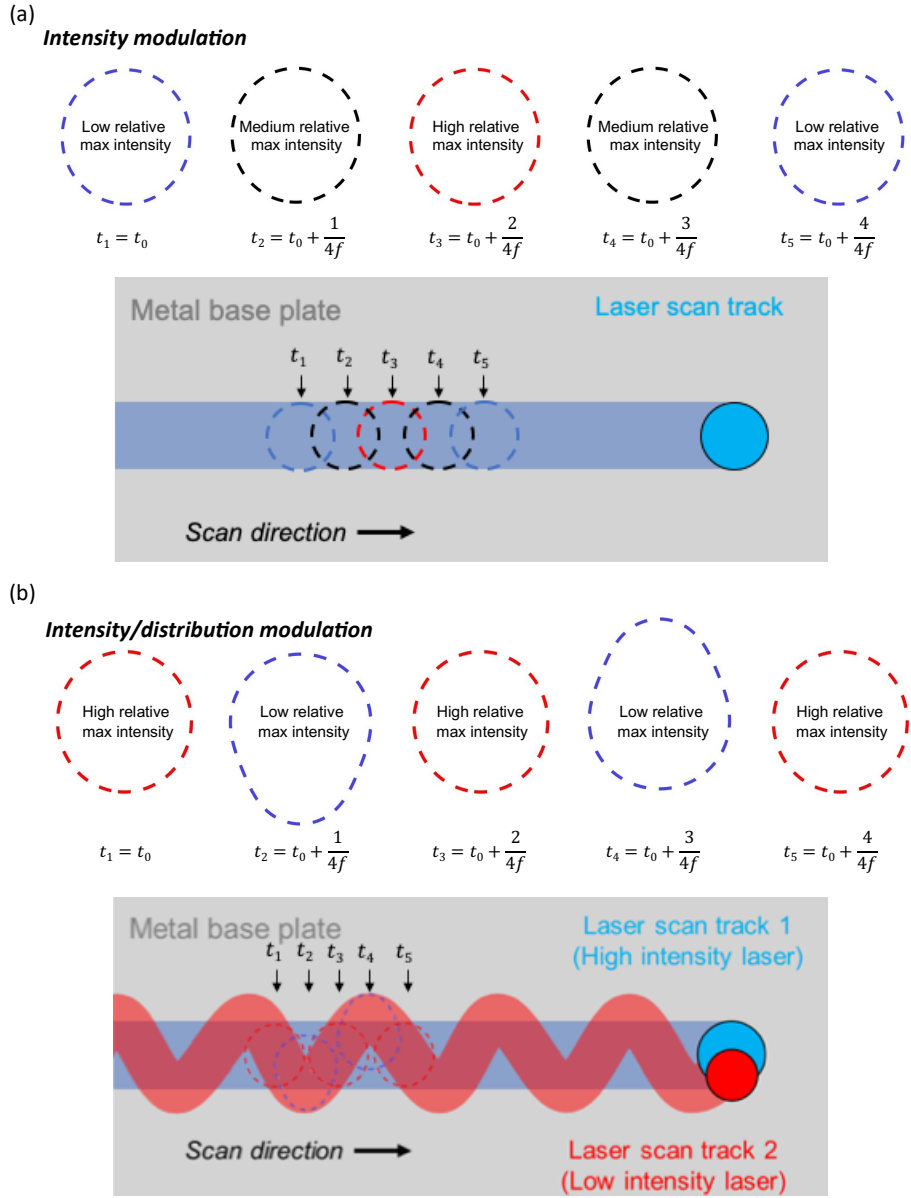

Figure 1: Illustrations showing beam shape contours throughout one modulation period and a supportive schematic showing the laser scanning schemes for (a) intensity modulation and (b) intensity distribution modulation. The colors of the beam shape contours denote the relative maximum beam intensity (red – high, black – medium, blue – low).

3. Results and Discussion

A CW laser beam with a sinusoidally modulated laser intensity (i.e., periodically fluctuating laser power) was scanned once across the surface of a metal AW5083 sheet to form a melt track on the surface. The frequency of laser intensity modulation (i.e., modulation frequency) was altered and the height profiles of the resulting melt track along the scan direction were analyzed through confocal microscopy (Fig. 2(a)). For a modulation frequency of 0 Hz (i.e., constant laser power), while random fluctuations in the height occur, there are no evident periodic surface features. On the other hand, for modulation frequencies of 100–1000 Hz, a clearly defined periodic pattern in the surface profile is observed. Moreover, a relationship between the modulation frequency and the height as well as the spatial frequency (i.e., distance between two consecutive surface features) of the periodic surface features can be identified. For example, the maximum height and spatial frequency are greater for a modulation frequency of 100 Hz compared to that of 1000 Hz, suggesting that the surface features formed for a modulation frequency of 100 Hz are more pronounced and further spaced.

To quantitatively characterize the periodic surface features, FFT analyses of the obtained height profiles were conducted (Fig. 2(b)). With no modulation of laser intensity (i.e., 0 Hz), no distinctive peaks are visible from the FFT spectra. On the other hand, with a modulation of laser intensity (i.e., 100–1000 Hz), distinctive peaks are present in the FFT spectra, indicating the existence of periodically orientated surface features. The measured spatial frequencies of the FFT peaks are approximately 100, 250, 500, and 1000 m$^{-1}$, for laser intensity modulation frequencies of 100, 250, 500, and 1000 Hz, respectively. The measured spatial frequency $f_s$ and the modulation frequency of laser intensity $f_m$ seems to be related to scan speed of the laser beam $v$ according to Eq.1.

$$f_s = \frac{f_m}{v} \qquad \text{EQ. 1}$$

Furthermore, the intensities of the peaks also depend highly on the modulation frequency. With an increase in modulation frequency from 100 Hz to 250 Hz, an increase in peak intensity is observed. However, as the modulation frequency is further increased to 500 Hz and 1000 Hz, a continuous decrease in peak intensity is observed. These ordered variations in peak intensities and the dominance of the 250 Hz peak suggests the presence of an underlying frequency coupling between the material and the modulation frequency, leading to more pronounced and ordered surface features.

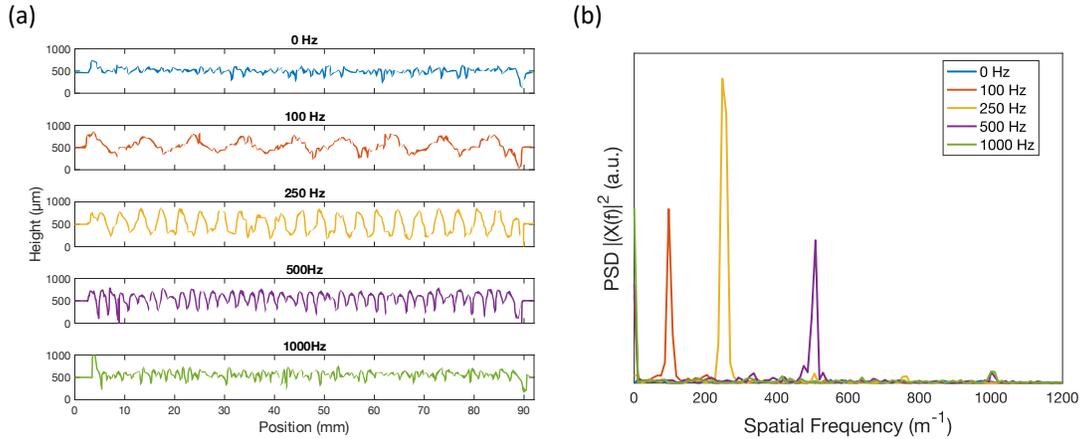

Figure 2: (a) Confocal microscopy height profiles obtained along the respective centers of the resulting melt tracks for sinusoidal intensity modulation frequencies of 0, 100, 250, 500, and 1000 Hz. (b) FFT spectra of the obtained height profiles for different modulation frequencies of laser intensity.

Next, the spatial distribution of laser intensity is sinusoidally modulated by the super-positioning of a spatially oscillating and non-oscillating laser beam (Fig. 1(b)). Similar to the laser intensity modulation experiments, the modulation frequency is altered, and the height profiles of the resulting melt tracks are obtained to conduct FFT analyses (Fig. 3). Consistent with Fig. 2(b), for no modulation ($f_m$ = 0 Hz), distinctive peaks cannot be confirmed, whereas for a modulation ($f_m$ > 0 Hz) distinctive peaks are confirmed from the obtained FFT spectra. The measured spatial frequencies of the surface features are each approximately 2000, 5000, 10000, and 20000 m$^{-1}$ for a modulation frequency of 100, 250, 500, 1000 Hz, respectively. In accordance with results obtained in the case of intensity modulation (Fig. 2(b)), the largest peak is observed for a modulation frequency of 250 Hz, and for higher modulation frequencies a decreasing trend can be observed. However, contrary to Fig. 2(b), the spatial frequency at which these peaks occur differ by a factor of 20. This is attributable to the fact that the scan speed was ten times smaller and maximum intensity is reached twice per periodic modulation for the current two-laser approach (Fig. 1(b)), resulting in spatial frequencies that are double the value expected according to Eq.1.

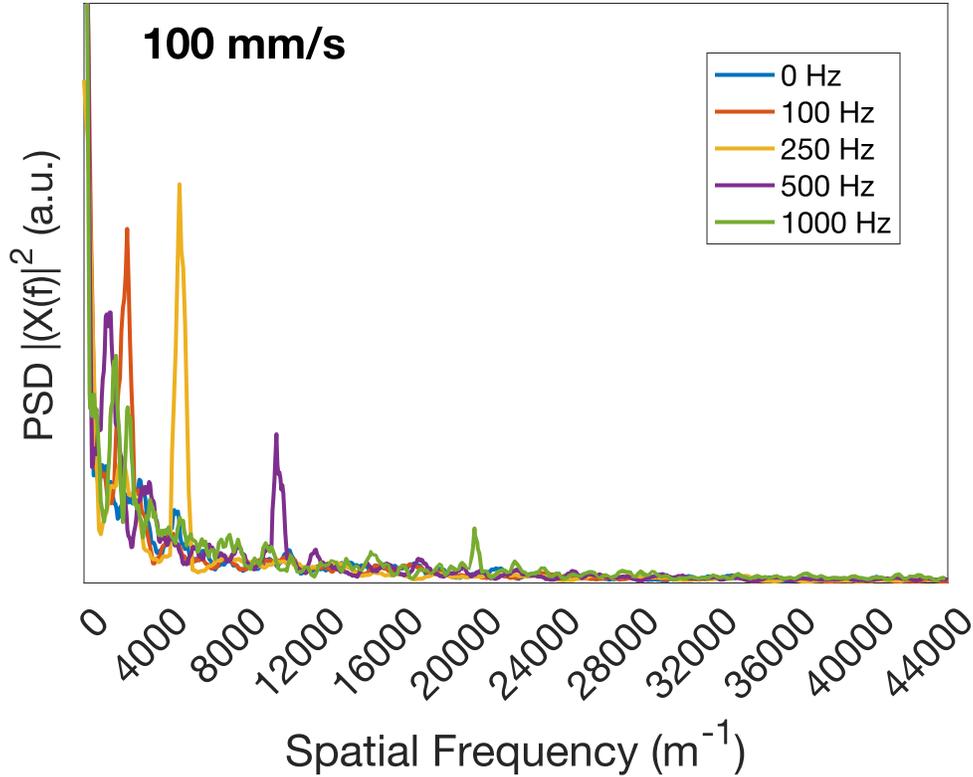

Figure 3: FFT spectra of the obtained height profiles for different modulation frequencies of laser intensity distribution.

To confirm the discussions regarding the FFT analyses, the surfaces of the melt tracks resulting from intensity distribution modulations were observed using SEM. Fig. 4 shows SEM images of melt tracks formed at a scan speed of 100 mm/s with a modulated frequency of (a) 0 Hz, (b) 200 Hz, and (c) 1000 Hz, respectively. As seen in Fig. 4(a), when a laser beam with no modulation frequency (0 Hz) is scanned, a melt track with no visible periodic surface features is formed. On the other hand, when a laser beam with a modulation frequency (250 Hz) is scanned, periodic surface features are apparent (regions indicated by the yellow arrows in Fig. 4(b)). The distance between the periodic features ($\alpha$) in Fig. 4(b) is approximately 250 μm. This distance agrees with the peak location seen in Fig. 3, as the spatial frequency is the inverse of the distance between the periodic surface features. For a significantly high modulation frequency (1000 Hz), such periodic surface features become less prominent (Fig. 4(c)), agreeing well with the decrease in peak intensities observed from the FFT spectra (Fig. 3).

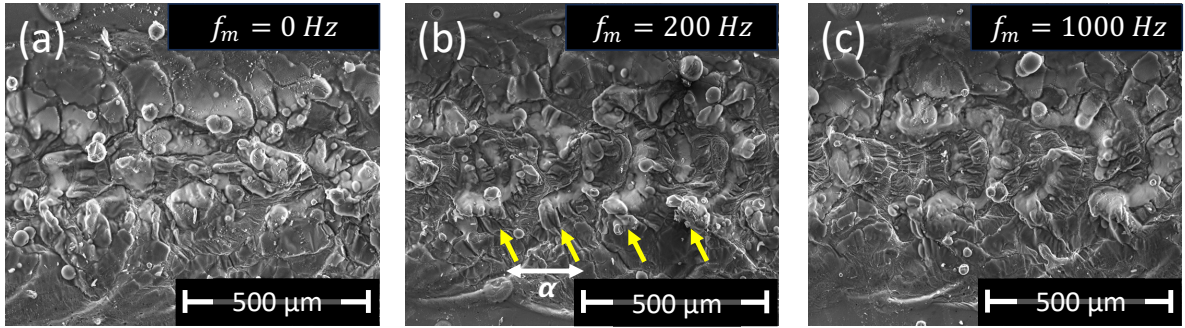

Figure 4: Surface SEM images of the melt track resulting from intensity distribution modulation frequencies of (a) 0 Hz, (b) 200 Hz, and (c) 1000 Hz. (b) displays periodic surface features (yellow arrows) and the distance between these periodic surface features is labeled as $\alpha$.

To further investigate the influence of the scan speed on these periodic surface features, the intensity distribution experiment was repeated at different scan speeds. Fig. 5 shows the obtained FFT spectra of the surface profiles for different modulation frequencies with scan speeds of (a) 50 mm/s, (b) 200 mm/s, and (c) 400 mm/s, respectively. Similar to the case of a scan speed of 100 mm/s (Fig. 3), distinctive peaks can be confirmed for a scanning speed of 50 mm/s (Fig. 5(a)). However, these peaks are located at different spatial frequencies, of approximately 4000, 10000, 20000, and 40000 m$^{-1}$, respectively. As the scanning speed is increased to 200 and 400 mm/s, the spatial frequencies at which peaks are observed shift to lower spatial frequencies (Fig. 5(b)). Moreover, the relative intensities of the peaks decrease with increasing scanning speed, suggesting less prominent surface features. This decrease in peak intensity may be attributable to a decrease in total energy input due to an increase in scanning speed, resulting in inadequate metal melting to form prominent periodic surface features (Fig. 5(c)). Nevertheless, consistent with the obtained FFT spectra shown thus far, the largest peak is observed for a modulation frequency of 250 Hz regardless of the scan speed.

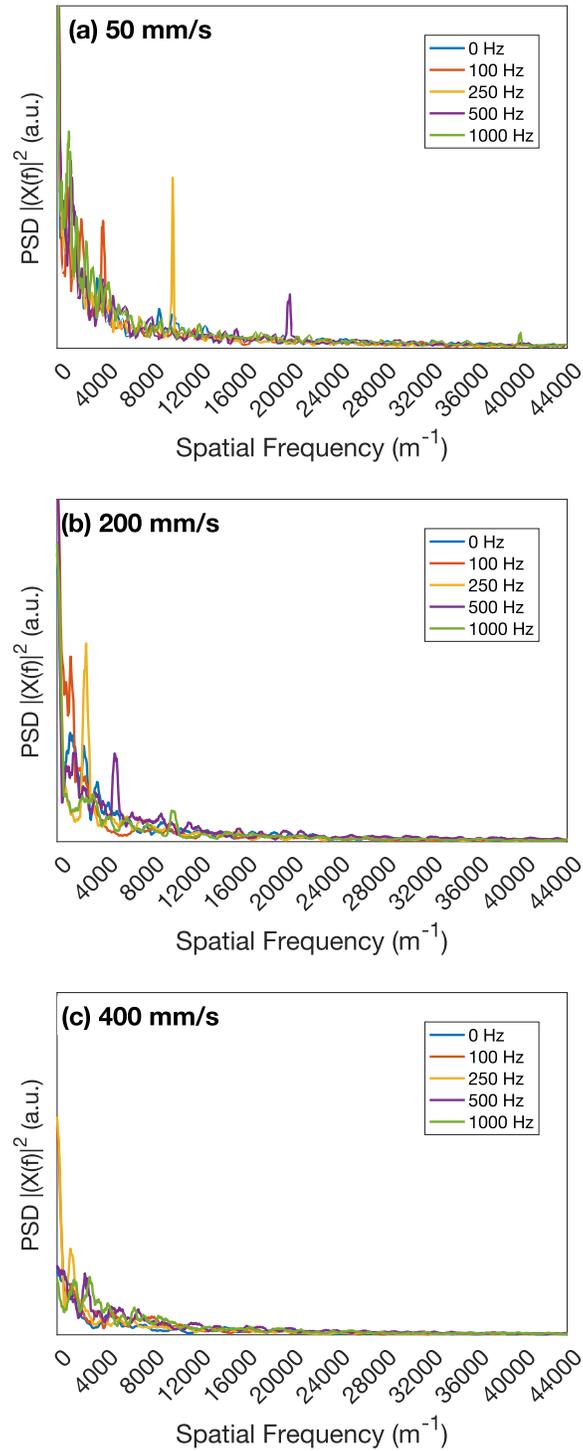

Figure 5: FFT spectra of the obtained height profiles for different modulation frequencies of laser intensity distribution, with a scan speed of (a) 50 mm/s, (b) 200 mm/s, and (c) 400 mm/s.

For all cases distinctive periodic patterns emerge when the laser beam intensity and distribution is modulated, contrasting random surface features generated by unmodulated intensities and intensity distributions. The FFT indicates that the spatial frequency at which these periodic surface features occur depend on the modulation frequency and the scan speed. Additionally, the FFT analyses of the melt track profiles reveal distinct peaks corresponding to modulation frequencies, with the most pronounced periodic features occurring at a specific modulation frequency, notably 250 Hz. This consistent association across all experiments implies a frequency coupling resonance at 250 Hz, independent of modulation type and scan speed. The exact reason of the underlying resonance cannot be concluded at this time, but its appearance can be attributed to an interplay between laser parameters and material-specific melt pool dynamics during melting and solidification phases. The melt pool can be coupled via both laser intensity modulation and laser intensity distribution, further contributing to the observed phenomena. Further investigation is necessary to understand the fundamental mechanisms underlying these laser material interactions and the significance of the observed frequency coupling resonance.

4. Conclusion

In this study, we investigated a laser driven resonance induced by dynamically modulating laser beam intensities and distributions during metal processing. When the laser intensity and/or spatial distribution was sinusoidally modulated, periodic surface features visibly different from those generated without modulation emerge within the melt track. The microscopy and FFT analyses conducted on the melt tracks clearly indicates that the spatial frequency at which these surface features occur, are highly dependent on the modulation frequency and the scan speed. Additionally, it was revealed that such periodic surface features become most pronounced for a specific modulation frequency. This was consistent whether the laser intensity, laser intensity distribution, or scanning speed was altered, implying the existence of a frequency coupling resonance, specifically 250 Hz in this study. The exact underlying mechanisms of such a coupling behavior cannot be concluded at this time, but such a phenomenon can be attributed to various factors, including spatial or temporal interferences of the laser-material interactions and/or inherent material-specific melt pool resonances. By further revealing the underlying mechanisms, the idea of modulating the laser intensity and/or intensity distribution may provide an important foundation to control and/or mitigation of inherent fluctuations in the melt pool through laser-driven resonances.

Acknowledgements

This work was funded by Princeton University, as well as by Deutsche Forschungsgemeinschaft (DFG, German Research Foundation) – 434946896 - within the framework of the Collaborative Research Centre SFB1120-236616214 "Bauteilpräzision durch Beherrschung von Schmelze und Erstarrung in Produktionsprozessen". A portion of the characterizations were conducted with the support of the Princeton Imaging and Analysis Center (IAC) which is partially supported by the Princeton Center for Complex Materials, a National Science Foundation (NSF)-MRSEC program (DMR-2011750). The experiments carried out at Beamline P07 of DESY PETRA III were in cooperation with Helmholtz-Zentrum Hereon in Hamburg as part of proposal BAG-20211050 and the authors would like to thank F. Beckmann, J. Moosmann, Christoph Spurk, Marc Hummel, Alexander Olowinsky, and all people involved for their support. The authors acknowledge the support of Cailabs for providing the CANUNDA setup. The authors thank Ankit Das for his insightful discussions.